\documentclass[a4]{article}

\newtheorem{proposition}{Proposition}
\newtheorem{lemma}{Lemma}
\newtheorem{corollary}{Corollary}
\newtheorem{theorem}{Theorem}

\def\Cntx{{\cal C}}

\def\cx#1{{\underline{#1}}}
\def\incx{\mathop{\triangleleft}}

\def\Real{{\bf R}}

\def\<{{\langle}}
\def\>{{\rangle}}
\def\||{{\parallel}}

\def\Hlbt{{\cal H}}
\def\PhaseSp{{\Omega}}
\def\Step#1{\noindent{\underline{\it Step#1.} \\}}

\def\Proof{\noindent{{\it Proof}\ .\/ }}
\def\Remark{{\noindent{\bf Remark}. \ }}
\def\QED{{\hfill \frame{\phantom{I}}}\\}
\def\span{\mathop{\rm span}}

\def\cket#1{{\left| #1 \right\rangle}}

\def\split#1{{C^{|\varphi\>; \epsilon}_{\cx{#1}}}}
\def\hsplit#1{{C^{|\varphi\>; \epsilon}_{  #1  }}}
\def\hSplit#1{{T_{  #1  }}}
\def\canonical#1#2{{{\tau}_{\cx{#1}\to\cx{#2}}}}
\def\Unitary#1#2{{U_{\cx{#1}\to\cx{#2}}}}
\def\uNitary#1#2{{U^{\dagger}_{\cx{#1}\to\cx{#2}}}}

\def\coord{\iota_{\cx{\alpha}}}
\def\Borel{{\cal B}}

\begin{document}
\title{Value Assignments to Observables Depending on a History of Context Change}

\author{Satoshi Uchiyama\thanks{email: uchiyama@hokusei.ac.jp}  \thanks{Address untill the end of Feb. 2007: Centro Vito Volterra, Universit\`a degli Studi di Roma
``Tor Vergata", 
Via Columbia, Rome 00133, Italy.}\\
Department of Life and Creative Sciences \\
Hokusei Gakuen University Junior College\\
Atsubetsu-ku, Sapporo 004-8631, Japan.\\
}


\maketitle


\begin{abstract}
The functional composition principle is generalized by taking into
account of history of context change. Analysis of Peres' example
shows hysteresis of value assignments. It is shown that value
assignments which depend on the history of context change are
possible in the case that the Hilbert space of state vectors is
finite dimensional.
\end{abstract}

\maketitle

\section{A generalized functional composition principle depending on history of context change}
The Kochen--Specker theorem states that there is no value assignment for quantum-mechanical observables that satisfies the functional composition principle (FUNC) when the dimension of the Hilbert space of state vectors is greater than two \cite{KochenSpecker, Redhead}.
By analogy with Riemannian surfaces which settle contradictive concept of multivalued functions by analytic continuation,
we investigate a value assignment depending on a history of context change.

The eigenvalues of an observable represented by a self-adjoint operator are values which are obtainable when one makes an observation.
If there was hidden variables that specify the observed value, we can assign the value to the pair of the observable and the hidden variables.
If we denote the hidden variables, the observable, and the eignevalue
 by $\omega$, $\hat{O}$, and $o_{i}$, respectively, then a  value assigning map $v$ is defined as a mapping of the set of all observables into a set of random variables such that for the possible $\omega$,
\begin{equation}
v(\hat{O})(\omega) = o_{i}.
\end{equation}

The functional composition principle (FUNC) states that for
observables $\hat{A}$ and $\hat{B}$, if there exists a function
$f$ such that $\hat{A} = f (\hat{B})$, then a value assigning map
$v$ satisfies that
\begin{equation}
v(\hat{A})(\omega) = f(v(\hat{B})(\omega)).
\end{equation}
Physically, a context of a measurement of an observable $\hat{O}$
is determined by a setting of the measurement apparatus.
Mathematically, a maximal Boolean sublattice of a lattice of
observational propositions on a quantum system determines a
context of a measurement. An observational proposition of an
yes-no experiment is represented by a projection operator. 
There is a one-to-one mapping of the set of all unit vectors into the set of projection operators of rank 1.
By using this mapping, we can specify a maximal Boolean sublattice, i.e., a context, by a complete orthonormal system (CONS) of vectors.
We define an equivalence relation on the set of all CONSs.
We shall say that two CONSs $(|\alpha_{1}\>, \ldots, |\alpha_{n}\> )$ and $(|\beta_{1}\>, \ldots, |\beta_{n}\> )$ are equivalent if there exist a permutation 
$p$ of the index set $\{1, 2, \ldots, n\}$ and phase factors $\exp(\sqrt{-1} \theta_{i})$s ($i=1, 2, \ldots, n$) such that $|\beta_{j}\> = \exp(\sqrt{-1} \theta_{p(j)})|\alpha_{p(j)}\>$ ($j=1, 2, \ldots, n$).
Thus a context is an equivalent class with respect to this equivalence relation and some CONS is a representative of a context. We denote a set of contexts
by $\Cntx$. $\Cntx$ is a subset of the quotient space with respect
to the equivalence relation of the complex Stiefel manifold of orthonormal
$n$-frames .

If a measurement apparatus that is not set in a context of a
measurement of an observable $\hat{O}$ gave outcomes, then results
would be invalid, i.e., the obtained values could be different
from the eigenvalues. 
On the contrary, a measurement apparatus that is set in the context of a measurement of the observable
$\hat{O}$ gives eigenvalues of $\hat{O}$ as outcomes. In this
sense, the measurement apparatus gives  stable results for
$\hat{O}$ in the context. We say that an observable $\hat{O}$ is
{\it stable in a context} $\cx{\alpha}\in \Cntx$ and denoting it
by $\hat{O}\incx \cx{\alpha}$, if $\hat{O}$ is diagonalized in a
CONS $(|\alpha_{1}\>, \ldots, |\alpha_{n}\> )$ that is a
representative of $\cx{\alpha}$.

When an observable $\hat{O}$ is not stable in a context $\cx{\alpha}$, there is no experimental restriction on values of $\hat{O}$.
Hence it is natural to consider that a value assigning map depends on contexts, and therefore the domain of a value assigning map is the set of observables that are stable in a given context.
Thus the contextual version of FUNC becomes as follows.

\noindent
{\bf Contextual version of FUNC (cFUNC)} \ If
$
\hat{A} = f(\hat{B})
$
and $\hat{B} \incx \cx{\alpha}$ then,
\begin{equation}
v_{\cx{\alpha}}(\hat{A}) = f(v_{\cx{\alpha}}(\hat{B})), \mbox{ a.s. on } \PhaseSp.
\end{equation}
Here, $\Omega$ is the set of hidden variables and a probability measure on $\Omega$ characterizing the  ensemble of the hidden varibles is supposed to be.

It is more general to consider that a value assigning map may depend on the contexts that the system experienced in the past.
A history of context change is a finite sequence of contexts $(\cx{\alpha}, \cx{\beta}, \ldots, \cx{\gamma})$, where $\cx{\alpha}, \cx{\beta}, \ldots, \cx{\gamma}\in \Cntx$.
We use a more intuitive notation about this such as $\cx{\alpha}\to \cx{\beta} \to \cdots \to \cx{\gamma}$.

\noindent
{\bf Generalized version of FUNC (gFUNC)} \ If
$
\hat{A} = f(\hat{B})
$
and $\hat{B} \incx \cx{\alpha}$ then,
\begin{equation}
v_{\cdots \to \cx{\beta} \to \cx{\alpha}}(\hat{A}) = f(v_{\cdots \to \cx{\beta} \to \cx{\alpha}}(\hat{B})), \mbox{ a.s. on } \PhaseSp
\end{equation}
for $\forall \cx{\beta}, \cdots \in \Cntx$,
where $\cdots \to$ represents the same history of context change in both sides of the equation.\\

\begin{proposition}
Provided  gFUNC, if observables $\hat{A}$ and $\hat{B}$ are stable in a context $\cx{\alpha}$, then
\begin{eqnarray}
v_{\cdots \to \cx{\alpha}}(\hat{A} + \hat{B}) &=&  v_{\cdots \to \cx{\alpha}}(\hat{A}) \ + \ v_{\cdots \to \cx{\alpha}}(\hat{B}),\\
v_{\cdots \to \cx{\alpha}}(\hat{A}\cdot \hat{B}) &=& v_{\cdots \to \cx{\alpha}}(\hat{A}) \ \cdot\  v_{\cdots \to \cx{\alpha}}(\hat{B}),
\mbox{ a.s. on } \Omega.
\end{eqnarray}
\end{proposition}
\Proof Since $\hat{A}$, $\hat{B}\incx \cx{\alpha}$, they can be
diagonalised simultaneously. Then there exist an observable
$\hat{C}$ and functions $g$ and $h$ such that $\hat{A} =
g(\hat{C})$ and $\hat{B} = h(\hat{C})$.
\begin{eqnarray*}
&& v_{\cdots \to \cx{\alpha}}(\hat{A} + \hat{B}) \\
&=&  v_{\cdots \to \cx{\alpha}}(g(\hat{C}) + h(\hat{C}))  \\
&=&  v_{\cdots \to \cx{\alpha}}( (g + h)( \hat{C}) )  \\
&\stackrel{gFUNC}{=}& (g + h) ( v_{\cdots \to \cx{\alpha}}(\hat{C}) )  \\
&=& g( v_{\cdots \to \cx{\alpha}}(\hat{C}) ) + h( v_{\cdots \to \cx{\alpha}}(\hat{C}) )  \\
&\stackrel{gFUNC}{=}& v_{\cdots \to \cx{\alpha}}( g(\hat{C}) ) +  v_{\cdots \to \cx{\alpha}}( h(\hat{C}) ) \\
&=&  v_{\cdots \to \cx{\alpha}}(\hat{A}) \ + \ v_{\cdots \to \cx{\alpha}}(\hat{B}).
\end{eqnarray*}
\begin{eqnarray*}
&& v_{\cdots \to \cx{\alpha}}(\hat{A} \cdot \hat{B}) \\
&=&  v_{\cdots \to \cx{\alpha}}(g(\hat{C}) \cdot h(\hat{C}))  \\
&=&  v_{\cdots \to \cx{\alpha}}( (g \cdot h)( \hat{C}) )  \\
&\stackrel{gFUNC}{=}& (g \cdot h) ( v_{\cdots \to \cx{\alpha}}(\hat{C}) )  \\
&=& g( v_{\cdots \to \cx{\alpha}}(\hat{C}) ) \cdot h( v_{\cdots \to \cx{\alpha}}(\hat{C}) )  \\
&\stackrel{gFUNC}{=}& v_{\cdots \to \cx{\alpha}}( g(\hat{C}) ) \cdot  v_{\cdots \to \cx{\alpha}}( h(\hat{C}) ) \\
&=&  v_{\cdots \to \cx{\alpha}}(\hat{A}) \ \cdot \ v_{\cdots \to \cx{\alpha}}(\hat{B}).
\end{eqnarray*}
\QED

\Remark Difficulty of defining a value assigning map with the FUNC appears when the $\hat{A}$ is
degenerated. Consider the following matrices:
$$
\begin{array}{c}
\hat{B} := \left(
\begin{array}{cc|c}
1 & 0 &  0 \\
0 & 2 &  0 \\
\hline
0 & 0 &  3
\end{array} \right),\\
\hat{C} := \left(
\begin{array}{rr|r}
\frac{1}{\sqrt{2}} & -\frac{1}{\sqrt{2}} &  0 \\
\frac{1}{\sqrt{2}} & \frac{1}{\sqrt{2}} &  0 \\
\hline
0 & 0 &  1
\end{array} \right)
\left(
\begin{array}{rr|r}
1 & 0 &  0 \\
0 & 2 &  0 \\
\hline
0 & 0 &  3
\end{array} \right)
\left(
\begin{array}{rr|r}
\frac{1}{\sqrt{2}} & \frac{1}{\sqrt{2}} &  0 \\
-\frac{1}{\sqrt{2}} & \frac{1}{\sqrt{2}} &  0 \\
\hline
0 & 0 &  1
\end{array} \right)
 = \left(
\begin{array}{rr|r}
\frac{3}{2} & -\frac{1}{2} &  0 \\
-\frac{1}{2} & \frac{3}{2} &  0 \\
\hline
0 & 0 &  3
\end{array} \right).
\end{array}
$$
Suppose  functions $f$ and $g$ are give by
\begin{equation}
f(x) = \left\{
\begin{array}{rl}
2 & \mbox{ if } x < 3\\
x & \mbox{ otherwise}
\end{array}\right., \quad
g(x) = \left\{
\begin{array}{rl}
2 & \mbox{ if } x \leq 2\\
x & \mbox{ otherwise}
\end{array}\right.,
\end{equation}
respectively.
Then
\begin{equation}
f(\hat{B}) =
\left(
\begin{array}{rr|r}
2 & 0 & 0\\
0 & 2 & 0\\
\hline
0 & 0 & 3
\end{array}\right).
\nonumber
\end{equation}
\begin{eqnarray*}
g(\hat{C}) &=&
\left(
\begin{array}{rr|r}
\frac{1}{\sqrt{2}} & -\frac{1}{\sqrt{2}} &  0 \\
\frac{1}{\sqrt{2}} & \frac{1}{\sqrt{2}} &  0 \\
\hline
0 & 0 &  1
\end{array} \right)
\left(
\begin{array}{rr|r}
2 & 0 &  0 \\
0 & 2 &  0 \\
\hline
0 & 0 &  3
\end{array} \right)
\left(
\begin{array}{rr|r}
\frac{1}{\sqrt{2}} & \frac{1}{\sqrt{2}} &  0 \\
-\frac{1}{\sqrt{2}} & \frac{1}{\sqrt{2}} &  0 \\
\hline
0 & 0 &  1
\end{array} \right) \\
&=&\left(
\begin{array}{rr|r}
2 & 0 & 0\\
0 & 2 & 0\\
\hline
0 & 0 & 3
\end{array}\right) = f(\hat{B}) = \hat{A}.
\end{eqnarray*}
Although $\hat{B}$ and $\hat{C}$ are stable in different contexts respectively,
$\hat{A}$ is stable in both of the contexts.
Thus one has to assign a value to $\hat{A}$ so that the assigned value would be consistent with values assigned to $\hat{B}$ and $\hat{C}$.
\\

Since cFUNC and gFUNC give no constraint among value assignments in different contexts,  the space of hidden variables is broken up into independent probability spaces of different contexts.
To synthesize them, we require the following condition.

\noindent
{\bf Non-transition condition (n-TRNS)}. \
For all contexts $\cx{\alpha}$, $\cx{\beta} \in \Cntx$, if $\hat{B} \incx \cx{\alpha}$ and $\hat{B}\incx \cx{\beta}$, then
\begin{equation}
v_{\cdots \to \cx{\alpha}}(\hat{B})(\omega) = v_{\cdots \to \cx{\alpha} \to \cx{\beta} }(\hat{B})(\omega), \mbox{ a.s. on } \PhaseSp,
\end{equation}
where $\cdots \to$ represents the same history of context change in both sides of the equation.\\

\Remark In the EPR--Bohm {\it Gedankenexperiment}, since the spin
observables of particles in a pair  commute, n-TRNS means that
an assigned value of a component of the spin of one particle does not
change,  when we change the setting of the measuring apparatus for
the spin of the other particle.
But locality in this sense is weaker than the genuine locality, since it allows  hysteresis of value assignments.\\

In the rest of this section, we clarify what kinds of observables
are stable in two different contexts.

\begin{proposition}\label{DEGENERATE}
There exists a nonzero self-adjoint operator that is diagonalized  in  two different CONSs $(|\alpha_{1}\>, \ldots, |\alpha_{n}\>)$ and $(|\beta_{1}\>, \ldots, |\beta_{n}\>)$
if and only if
there exist partitions $\{I_{1}, \ldots, I_{m}\}$ and $\{ J_{1}, \ldots, J_{m} \}$ of the index set $X := \{1, \ldots, n\}$  ($m \leq n$) such that
\begin{equation}
\span \left\{ |\alpha_{i}\> : i \in I_{k} \right\} = \span \left\{ |\beta_{j}\> : j \in J_{k} \right\},
\end{equation}
for $k = 1, \ldots, m$.
\end{proposition}
\Proof ($\Rightarrow$) Let $\hat{O}$ be a self-adjoint operator
that is diagonalized  in both  CONSs $(|\alpha_{1}\>, \ldots,
|\alpha_{n}\>)$ and $(|\beta_{1}\>, \ldots, |\beta_{n}\>)$.
Suppose $\hat{O}$ has eigenvalues $o_{1}, \ldots , o_{m}$ that are
distinct from each other ($m \leq n$). We define a partition $\{
I_{k} \}$ of the index set $X$ by
\begin{equation}
I_{k} : = \left\{ i \in X  :\  \hat{O} |\alpha_{i}\> = o_{k}|\alpha_{i}\> \right\}.
\end{equation}
Similarly, we define  a partition $\{ J_{k} \}$ of $X$ by
\begin{equation}
J_{k} : = \left\{ j \in X  :\  \hat{O} |\beta_{j}\> = o_{k}|\beta_{j}\> \right\}.
\end{equation}
Since $\span \{ |\alpha_{i}\> : i \in I_{k} \}$ and $\span \{ |\beta_{i}\> : i \in J_{k} \}$ is the eigenspace of $\hat{O}$ with the eigenvalue $o_{k}$, they coincide with each other.

($\Leftarrow$)
Let $o_{k}$s are increasing real numbers, i.e., $o_{1} < o_{2} < \ldots < o_{m}$.
We define a self-adjoint operator $\hat{O}$ by
\begin{equation}
\hat{O} := \sum_{k = 1}^{m} o_{k} \sum_{i \in I_{k}} |\alpha_{i}\>\<\alpha_{i}|;
\end{equation}
this is diagonalized in $(|\alpha_{1}\>, \ldots, |\alpha_{n}\>)$.
Since $\sum_{i \in I_{k}} |\alpha_{i}\>\<\alpha_{i}| = \sum_{j \in J_{k}} |\beta_{j}\>\<\beta_{j}|$, $\hat{O}$ is diagonalized in $(|\beta_{1}\>, \ldots, |\beta_{n}\>)$.
\QED

\Remark
When $m=1$, the self-adjoint operator is $I$, i.e., the identity operator.
$m=n$ means that the two CONSs are representatives of the same context.\\

\begin{lemma}\label{FINEST}
For two different CONSs $(|\alpha_{1}\>, \ldots, |\alpha_{n}\>)$ and $(|\beta_{1}\>, \ldots, |\beta_{n}\>)$, there exists  uniquely the finest partitions $\{I_{1}, \ldots, I_{m}\}$ and $\{ J_{1}, \ldots, J_{m} \}$ of the index set $X := \{1, \ldots, n\}$ of the CONSs respectively ($m \leq n$) such that
\begin{equation}
\span \left\{ |\alpha_{i}\> : i \in I_{k} \right\} = \span \left\{ |\beta_{j}\> : j \in J_{k} \right\}, \label{PARTITION}
\end{equation}
for $k = 1, \ldots, m$.
\end{lemma}
\Proof
Consider  partitions $\{I'_{1}, \ldots, I'_{m'}\}$ and $\{ J'_{1}, \ldots, J'_{m'} \}$ of the index set $X$ ($m' \leq n$) such that
\[
\span \left\{ |\alpha_{i}\> : i \in I'_{k} \right\} = \span \left\{ |\beta_{j}\> : j \in J'_{k} \right\},
\]
for $k = 1, \ldots, m'$.
If there exist another  partitions $\{I''_{1}, \ldots, I''_{m''}\}$ and $\{ J''_{1}, \ldots, J''_{m''} \}$ of the index set $X$  ($m'' \leq n$) such that
\[
\span \left\{ |\alpha_{i}\> : i \in I''_{k} \right\} = \span \left\{ |\beta_{j}\> : j \in J''_{k} \right\},
\]
for $k = 1, \ldots, m''$, then put
\begin{eqnarray*}
\tilde{I}_{(k, l)} &:=& I'_{k} \cap I''_{l},\\
\tilde{J}_{(k, l)} &:=& J'_{k} \cap J''_{l},
\end{eqnarray*}
for $k = 1, \ldots , m'$, $l=1, \ldots, m''$.
Since
\[
\span \left\{ |\alpha_{i}\> :\ i\in \tilde{I}_{(k, l)} \right\}
= \span \left\{ |\alpha_{i}\> :\ i\in I'_{k} \right\}
\cap \span \left\{ |\alpha_{i}\> :\ i\in I''_{l} \right\}
\]
and
\[
\span \left\{ |\beta_{j}\> :\ j\in \tilde{J}_{(k, l)} \right\}
= \span \left\{ |\beta_{j}\> :\ j\in J'_{k} \right\}
\cap \span \left\{ |\beta_{j}\> :\ i\in J''_{l} \right\},
\]
we obtain
\begin{equation}
\span \left\{ |\alpha_{i}\> :\ i\in \tilde{I}_{(k, l)} \right\}
=
\span \left\{ |\beta_{j}\> :\ j\in \tilde{J}_{(k, l)} \right\}. \label{EqualSpan}
\end{equation}
Thus $\{\tilde{I}_{(k, l)}\}$ and $\{\tilde{J}_{(k, l)}\}$ are
finer partitions  of $X$ satisfying the condition
(\ref{EqualSpan}) than $\{I'_{k}\}$ and $\{J'_{l}\}$ or coincide with $\{I'_{k}\}$ and $\{J'_{l}\}$. 
Since the
index set $X$ is finite, above procedure ends with finite times
and we obtain the finest partitions $\{I_{i} \}$ and $\{J_{i} \}$
satisfying the condition (\ref{PARTITION}). The uniqueness follows
from the maximality of the $\{I_{i} \}$ and the $\{J_{i} \}$. \QED

\begin{corollary}\label{SPECDEC}
For two different contexts $\cx{\alpha} \equiv (|\alpha_{1}\>, \ldots, |\alpha_{n}\>)$ and $\cx{\beta} \equiv (|\beta_{1}\>, \ldots, |\beta_{n}\>)$,
an observable $\hat{O}$ is stable in $\cx{\alpha}$ and  $\cx{\beta}$
if and only if
 $\hat{O}$ can be written as
\begin{equation}
\hat{O} = \sum_{k = 1}^{m} o_{k} P_k,
\end{equation}
where $o_{k}$ is an eigenvalue of $\hat{O}$ and $P_k$ is given as
$
P_k := \sum_{i\in I_{k}} |\alpha_{i}\>\<\alpha_{i}| = \sum_{j \in J_{k}} |\beta_{j}\>\<\beta_{j}|
$
by using the finest partitions $\{I_{1}, \ldots,  I_{m}\}$ and the $\{J_{1}, \ldots, J_{m} \}$ of the index set given by Lemma \ref{FINEST}.
\end{corollary}
\Proof
It follows from the spectral decomposition of $\hat{O}$, Proposition \ref{DEGENERATE} and Lemma \ref{FINEST}.
\QED

\section{Contextuality in the Peres' example of nonexistence of noncontextual hidden variables}
Peres showed that it is impossible to assign a value to $(\sigma_{x}\otimes\sigma_{y})\cdot(\sigma_{y}\otimes\sigma_{x})$ noncontextually in the singlet state \cite{Peres}.
We denote the eigenvectors with eigenvalue $\pm 1$ of $\sigma_{X}$ by $|X\pm\>$ ($X = x, y, z$).
Here are the contexts that appear in the Peres' example:
\begin{eqnarray}
\cx{\alpha} &\equiv& \left(\ |x+\>\otimes|x+\>,\  |x-\>\otimes|x-\>,\ |x+\>\otimes|x-\>,\ |x-\>\otimes|x+\>\       \right), \label{CntxX} \\
\cx{\beta}  &\equiv& \left(\ |y+\>\otimes|y+\>,\  |y-\>\otimes|y-\>,\ |y+\>\otimes|y-\>,\ |y-\>\otimes|y+\>\       \right), \label{CntxY} \\
\cx{\gamma}  &\equiv& \left(\ |z+\>\otimes|z+\>,\  |z-\>\otimes|z-\>,\ |z+\>\otimes|z-\>,\ |z-\>\otimes|z+\>\       \right),  \label{CntxZ}\\
\cx{\delta} &\equiv& \left(\ |x+\>\otimes|y+\>,\  |x-\>\otimes|y-\>,\ |x+\>\otimes|y-\>,\ |x-\>\otimes|y+\>\       \right), \nonumber \\
\cx{\epsilon} &\equiv& \left(\ |y+\>\otimes|x+\>,\  |y-\>\otimes|x-\>,\ |y+\>\otimes|x-\>,\ |y-\>\otimes|x+\>\       \right), \nonumber \\
\cx{\xi} &\equiv& \left( \frac{1}{\sqrt{2}}(|z+\>\otimes|z+\> + \sqrt{-1}|z-\>\otimes|z-\> ), \right. \nonumber \\
& & \frac{1}{\sqrt{2}}(|z+\>\otimes|z+\> - \sqrt{-1}|z-\>\otimes|z-\> ),  \nonumber \\
& & \ \  \frac{1}{\sqrt{2}}(|z+\>\otimes|z-\> + \sqrt{-1}|z-\>\otimes|z+\> ), \nonumber \\
& & \left. \frac{1}{\sqrt{2}}(|z+\>\otimes|z-\> - \sqrt{-1}|z-\>\otimes|z+\> ) \right).
\nonumber
\end{eqnarray}
We can see the following:
\begin{eqnarray*}
\cx{\alpha} &\triangleright&  \sigma_x \otimes I,\   I \otimes \sigma_x,\  \sigma_x\otimes \sigma_x, \\
\cx{\beta} &\triangleright&  \sigma_y \otimes I,\   I \otimes \sigma_y, \  \sigma_y\otimes \sigma_y, \\
\cx{\gamma} &\triangleright&  \sigma_z \otimes I,  \ I \otimes \sigma_z, \  \sigma_z\otimes \sigma_z, \\
\cx{\delta} &\triangleright&  \sigma_x \otimes I, \ I \otimes \sigma_y, \  \sigma_x\otimes \sigma_y,\\
\cx{\epsilon} &\triangleright&  \sigma_y \otimes I,\ I \otimes \sigma_x, \  \sigma_y\otimes \sigma_x,\\
\cx{\xi} &\triangleright&   \sigma_x \otimes\sigma_y, \ \sigma_y\otimes \sigma_x, \ \sigma_z\otimes \sigma_z.
\end{eqnarray*}

We analyze the Peres' example by exploiting gFUNC and n-TRNS. In order
to do this, we need one more condition.

\noindent
{\bf Perfect anti-correlation (a-CRL)}. \
For the contexts $\cx{\alpha}$ of (\ref{CntxX}), $\cx{\beta}$ of (\ref{CntxY}), $\cx{\gamma}$ of (\ref{CntxZ}),
\begin{eqnarray*}
v_{\cdots \to \cx{\alpha}}(\sigma_x\otimes I)(\omega) &=& - v_{\cdots \to \cx{\alpha} }( I\otimes \sigma_x)(\omega), \\
v_{\cdots \to \cx{\beta}}(\sigma_y\otimes I)(\omega) &=& - v_{\cdots \to \cx{\beta} }( I\otimes \sigma_y)(\omega), \\
v_{\cdots \to \cx{\gamma}}(\sigma_z\otimes I)(\omega) &=& - v_{\cdots \to \cx{\gamma} }( I\otimes \sigma_z)(\omega),
\end{eqnarray*}
respectively, for  all possible $\omega \in \Omega$.
Here $\cdots \to$ represents the same history of context change in  both sides of the equations.\\

\Remark
If we add a rule that we can replace an observable by its eigenvalue in the corresponding eigenstate, then a-CRL follows from gFUNC.\\

\begin{proposition}\label{PERES1}
If gFUNC, n-TRNS, a-CRL hold and eigenvalues are assigned  to the
corresponding observables, then
\begin{equation}
v_{\cx{\xi}}( (\sigma_x\otimes \sigma_y)\cdot(\sigma_y\otimes\sigma_x))
=- 1, \ \mbox{ a.s. } \Omega. \label{CntxXi}
\end{equation}
\end{proposition}

\Proof
\begin{eqnarray}
&&v_{\cx{\xi}}( (\sigma_x\otimes \sigma_y)\cdot(\sigma_y\otimes\sigma_x))  \nonumber\\
&=&v_{\cx{\xi}}( \sigma_z\otimes \sigma_z )  \nonumber\\
&\stackrel{n-TRNS}{=}&v_{\cx{\xi}\to\cx{\gamma}}( \sigma_z\otimes \sigma_z )  \nonumber\\
&=&v_{\cx{\xi}\to\cx{\gamma}}( (\sigma_z\otimes I)\cdot( I \otimes\sigma_z))  \nonumber\\
&\stackrel{gFUNC}{=}&v_{\cx{\xi}\to\cx{\gamma}}(\sigma_z\otimes I) \  v_{\cx{\xi}\to\cx{\gamma}}( I \otimes\sigma_z )  \nonumber\\
&\stackrel{a-CRL}{=}& - (v_{\cx{\xi}\to\cx{\gamma}}(\sigma_z\otimes I))^2 = -1.
\end{eqnarray}
The last equality follows from that the eigenvalues of $\sigma_z\otimes I$ are $\pm 1$.
\QED

\begin{proposition}\label{PERES2}
If gFUNC, n-TRNS, a-CRL hold, then
\begin{eqnarray}
&&v_{\cx{\xi}}( (\sigma_x\otimes \sigma_y)\cdot(\sigma_y\otimes\sigma_x)) \nonumber \\
&=&
v_{\cx{\xi}\to\cx{\delta} \to \cx{\alpha}}(\sigma_x\otimes I) \  v_{\cx{\xi}\to\cx{\delta}\to\cx{\beta}}(\sigma_y\otimes I) \ v_{\cx{\xi}\to\cx{\epsilon}\to\cx{\beta}}(\sigma_y\otimes I) \ v_{\cx{\xi}\to\cx{\epsilon}\to\cx{\alpha}}(\sigma_x \otimes I),    \label{Hist}
\end{eqnarray}
a.s. on  $\Omega$.
\end{proposition}

\Proof
\begin{eqnarray*}
&&v_{\cx{\xi}}( (\sigma_x\otimes \sigma_y)\cdot(\sigma_y\otimes\sigma_x))\\
&\stackrel{{gFUNC}}{=}& v_{\cx{\xi}}( \sigma_x\otimes \sigma_y ) \ v_{\cx{\xi}}( \sigma_y\otimes\sigma_x )\\
&\stackrel{n-TRNS}{=}& v_{\cx{\xi}\to\cx{\delta}}( \sigma_x\otimes \sigma_y ) \ v_{\cx{\xi}}( \sigma_y\otimes\sigma_x )\\
&{=}& v_{\cx{\xi}\to\cx{\delta}}( (\sigma_x\otimes I)(I\otimes \sigma_y)) \ v_{\cx{\xi}}(\sigma_y\otimes\sigma_x)\\
&\stackrel{gFUNC}{=}& v_{\cx{\xi}\to\cx{\delta}}(\sigma_x\otimes I) \  v_{\cx{\xi}\to\cx{\delta}}(I\otimes \sigma_y) \ v_{\cx{\xi}}(\sigma_y\otimes\sigma_x)\\
&\stackrel{n-TRNS}{=}& v_{\cx{\xi}\to\cx{\delta} \to \cx{\alpha}}(\sigma_x\otimes I) \  v_{\cx{\xi}\to\cx{\delta}}(I\otimes \sigma_y) \ v_{\cx{\xi}}(\sigma_y\otimes\sigma_x)\\
&\stackrel{n-TRNS}{=}& v_{\cx{\xi}\to\cx{\delta} \to \cx{\alpha}}(\sigma_x\otimes I) \  v_{\cx{\xi}\to\cx{\delta}\to\cx{\beta}}(I\otimes \sigma_y) \ v_{\cx{\xi}}(\sigma_y\otimes\sigma_x)\\
&\stackrel{a-CRL}{=}& - v_{\cx{\xi}\to\cx{\delta} \to \cx{\alpha}}(\sigma_x\otimes I) \  v_{\cx{\xi}\to\cx{\delta}\to\cx{\beta}}(\sigma_y\otimes I) \ v_{\cx{\xi}}(\sigma_y\otimes\sigma_x).
\end{eqnarray*}
We can proceed along calculatiing.
\begin{eqnarray*}
&&v_{\cx{\xi}}( (\sigma_x\otimes \sigma_y)\cdot(\sigma_y\otimes\sigma_x)) \nonumber\\
&\stackrel{n-TRNS}{=}& - v_{\cx{\xi}\to\cx{\delta} \to \cx{\alpha}}(\sigma_x\otimes I) \  v_{\cx{\xi}\to\cx{\delta}\to\cx{\beta}}(\sigma_y\otimes I) \ v_{\cx{\xi}\to\cx{\epsilon}}(\sigma_y\otimes\sigma_x) \nonumber\\
&\stackrel{gFUNC}{=}& - v_{\cx{\xi}\to\cx{\delta} \to \cx{\alpha}}(\sigma_x\otimes I) \  v_{\cx{\xi}\to\cx{\delta}\to\cx{\beta}}(\sigma_y\otimes I) \ v_{\cx{\xi}\to\cx{\epsilon}}(\sigma_y\otimes I) \ v_{\cx{\xi}\to\cx{\epsilon}}(I \otimes\sigma_x) \nonumber\\
&\stackrel{n-TRNS}{=}& - v_{\cx{\xi}\to\cx{\delta} \to \cx{\alpha}}(\sigma_x\otimes I) \  v_{\cx{\xi}\to\cx{\delta}\to\cx{\beta}}(\sigma_y\otimes I) \ v_{\cx{\xi}\to\cx{\epsilon}}(\sigma_y\otimes I) \ v_{\cx{\xi}\to\cx{\epsilon}\to\cx{\alpha}}(I \otimes\sigma_x) \nonumber\\
&\stackrel{a-CRL}{=}& v_{\cx{\xi}\to\cx{\delta} \to \cx{\alpha}}(\sigma_x\otimes I) \  v_{\cx{\xi}\to\cx{\delta}\to\cx{\beta}}(\sigma_y\otimes I) \ v_{\cx{\xi}\to\cx{\epsilon}}(\sigma_y\otimes I) \ v_{\cx{\xi}\to\cx{\epsilon}\to\cx{\alpha}}(\sigma_x \otimes I) \nonumber\\
&\stackrel{n-TRNS}{=}& v_{\cx{\xi}\to\cx{\delta} \to \cx{\alpha}}(\sigma_x\otimes I) \  v_{\cx{\xi}\to\cx{\delta}\to\cx{\beta}}(\sigma_y\otimes I) \ v_{\cx{\xi}\to\cx{\epsilon}\to\cx{\beta}}(\sigma_y\otimes I) \ v_{\cx{\xi}\to\cx{\epsilon}\to\cx{\alpha}}(\sigma_x \otimes I).
\end{eqnarray*}
\QED

\begin{proposition}
If gFUNC, n-TRNS, a-CRL hold, then a value assignment has hysteresis.
\end{proposition}
\Proof
By Proposition \ref{PERES1}, the r.h.s. of ($\ref{Hist}$) is equal to $-1$.
Because every terms of the r.h.s. of ($\ref{Hist}$) assume only the values $\pm1$,  exactly one of $v_{\cx{\xi}\to\cx{\delta} \to \cx{\alpha}}(\sigma_x\otimes I)$,  $v_{\cx{\xi}\to\cx{\delta}\to\cx{\beta}}(\sigma_y\otimes I)$,  $v_{\cx{\xi}\to\cx{\epsilon}\to\cx{\beta}}(\sigma_y\otimes I)$,  $v_{\cx{\xi}\to\cx{\epsilon}\to\cx{\alpha}}(\sigma_x \otimes I)$ must have opposite sign from the ones of the rests.
We note that $v_{\cx{\xi}\to\cx{\delta} \to \cx{\alpha}}(\sigma_x\otimes I) = v_{\cx{\xi}\to\cx{\epsilon}\to\cx{\alpha}}(\sigma_x \otimes I)$ and $v_{\cx{\xi}\to\cx{\delta}\to\cx{\beta}}(\sigma_y\otimes I)=v_{\cx{\xi}\to\cx{\epsilon}\to\cx{\beta}}(\sigma_y\otimes I)$ do not follow from n-TRNAS, because $\sigma_x\otimes I$ is not stable in the context $\cx{\epsilon}$ and $\sigma_y\otimes I$ is not stable in the context $\cx{\delta}$.
Therefore we obtain
\begin{equation}
v_{\cx{\xi}\to\cx{\delta} \to \cx{\alpha}}(\sigma_x\otimes I)=
v_{\cx{\xi}\to\cx{\epsilon}\to\cx{\alpha}}(\sigma_x \otimes I), \qquad
v_{\cx{\xi}\to\cx{\delta}\to\cx{\beta}}(\sigma_y\otimes I) \not=
v_{\cx{\xi}\to\cx{\epsilon}\to\cx{\beta}}(\sigma_y\otimes I),
\end{equation}
or
\begin{equation}
v_{\cx{\xi}\to\cx{\delta} \to \cx{\alpha}}(\sigma_x\otimes I)\not=
v_{\cx{\xi}\to\cx{\epsilon}\to\cx{\alpha}}(\sigma_x \otimes I), \qquad
v_{\cx{\xi}\to\cx{\delta}\to\cx{\beta}}(\sigma_y\otimes I) =
v_{\cx{\xi}\to\cx{\epsilon}\to\cx{\beta}}(\sigma_y\otimes I).
\end{equation}
This shows  hysteresis of the value assignment to $\sigma_x\otimes I$ or $\sigma_y\otimes I$.
\QED
\\

\Remark
$v_{\cx{\xi}\to\cx{\delta}\to\cx{\beta}}(\sigma_y\otimes I)$ shows that the system experienced the context $\cx{\delta}$ in which $\sigma_y\otimes I$ was not stable.
$v_{\cx{\xi}\to\cx{\epsilon}\to\cx{\alpha}}(\sigma_x \otimes I)$ shows that the system experienced the context $\cx{\epsilon}$ in which $\sigma_x\otimes I$ was not stable.
Only one of 
$v_{\cx{\xi}\to\cx{\delta}\to\cx{\beta}}(\sigma_y\otimes I)$ and $v_{\cx{\xi}\to\cx{\epsilon}\to\cx{\alpha}}(\sigma_x \otimes I)$, however, can assume values different from the values through a different values through a different history of context change.
This is  similar to the case of the Kocken--Specker paradox on spin-1 observables \cite{KochenSpecker}.
\\

\begin{proposition}
If gFUNC, n-TRNS, a-CRL hold and eigenvalues are assigned  to the
corresponding observables, then there is no noncontextual value
assigning map in the EPR--Bohm {\it Gedankenexperiment}.
\end{proposition}

\Proof
Suppose that the value assigning map is noncontextual.
Then (\ref{Hist}) becomes
\begin{eqnarray*}
&& v_{\cx{\xi}\to\cx{\delta} \to \cx{\alpha}}(\sigma_x\otimes I) \  v_{\cx{\xi}\to\cx{\delta}\to\cx{\beta}}(\sigma_y\otimes I) \ v_{\cx{\xi}\to\cx{\epsilon}\to\cx{\beta}}(\sigma_y\otimes I) \ v_{\cx{\xi}\to\cx{\epsilon}\to\cx{\alpha}}(\sigma_x \otimes I)\\
&=&  v(\sigma_x\otimes I) \  v(\sigma_y\otimes I) \ v(\sigma_y\otimes I) \ v(\sigma_x \otimes I)\\
&=&  v(\sigma_x\otimes I)^2 \  v(\sigma_y\otimes I)^2 = +1.
\end{eqnarray*}
Thus this contradicts (\ref{CntxXi}); there is no such value assigning map.
\QED

\Remark
This is nothing else but what Peres shows in ref. \cite{Peres}.\\


\section{Context changing maps}

We associate a mapping $\canonical{\alpha}{\beta}$ of $\PhaseSp$ into $\PhaseSp$ with a context change from a context $\cx{\alpha} \equiv  (|\alpha_{1}\>, \ldots, |\alpha_{n}\>) $ to a context $\cx{\beta} \equiv  (|\beta_{1}\>, \ldots, |\beta_{n}\>)$.

Let $\{ I_{k} \}$ and $\{ J_{k} \}$ be the finest partitions of the index set $\{1, \ldots, n\}$ of Lemma \ref{FINEST}.
Let $q$ be a permutation of the index set  such that $i \in I_{k}$ implies $q(i) \in J_{k}$ ($k = 1, \ldots, m$).
We define a unitary transformation $\Unitary{\alpha}{\beta}$ by
\begin{equation}
\Unitary{\alpha}{\beta} := \sum_{k=1}^{m} \sum_{i \in I_{k}} |\beta_{q(i)}\>\<\alpha_{i}|.
\end{equation}

$\Unitary{\alpha}{\beta}$ changes the context $\cx{\alpha}$ to $\cx{\beta}$.
Symbolically, we write
\begin{equation}
\cx{\beta} = \Unitary{\alpha}{\beta} \ \cx{\alpha}.
\end{equation}

\begin{proposition}\label{INVARIANT}
If $\hat{O} \incx \cx{\alpha}, \cx{\beta}$, then $\Unitary{\alpha}{\beta}\ \hat{O}\ \uNitary{\alpha}{\beta} = \hat{O}$.
\end{proposition}
\Proof
By Corollary \ref{SPECDEC} and the definition of
$\Unitary{\alpha}{\beta}$, denoting eigenvalues of $\hat{O}$ by
$o_{k}$s, we have
\begin{eqnarray*}
\Unitary{\alpha}{\beta}\ \hat{O}\ \uNitary{\alpha}{\beta} &=& \sum_{k}\sum_{i\in I_{k}} |\beta_{q(i)}\>\<\alpha_{i}| \sum_{l}o_{l} \sum_{i'\in I_{l}} |\alpha_{i'}\>\<\alpha_{i'}| \ \uNitary{\alpha}{\beta} \\
&=& \sum_{k}\sum_{i\in I_{k}} |\beta_{q(i)}\>o_{k}\<\alpha_{i}| \ \uNitary{\alpha}{\beta} \\
&=& \sum_{k}\sum_{i\in I_{k}} o_{k} |\beta_{q(i)}\>\<\alpha_{i}|  \sum_{l}  \sum_{i'\in I_{l}} |\alpha_{i'}\>\<\beta_{q(i')}| \\
&=& \sum_{k}\sum_{i\in I_{k}} o_{k} |\beta_{q(i)}\>\<\beta_{q(i)}| \\
&=& \sum_{k}\sum_{j\in J_{k}} o_{k} |\beta_{j}\>\<\beta_{j}| \\
&=& \hat{O}.
\end{eqnarray*}
\QED

We say that a value assigning map $ v_{\cdots\to\cx{\alpha}}$ {\it admits a context changing map} from $\cx{\alpha}$ to $\cx{\beta}$, if
for an arbitrary observable $\hat{O} \incx \cx{\alpha}$, there exists a one-to-one mapping $\canonical{\alpha}{\beta}$ of $\Omega$ onto $\Omega$ such that
\begin{equation}
v_{\cdots\to\cx{\alpha}\to\cx{\beta}}( \Unitary{\alpha}{\beta} \ \hat{O} \ \uNitary{\alpha}{\beta} )(\omega) = v_{\cdots\to\cx{\alpha}}(\hat{O})( \canonical{\alpha}{\beta}^{-1}(\omega) ), \qquad \mbox{ a.s. on } \PhaseSp.    \label{CONTCHG}
\end{equation}
We call such $\canonical{\alpha}{\beta}$ a {\it context changing map} from $\cx{\alpha}$ to $\cx{\beta}$.

\begin{proposition}\label{STABLE}
If a value assigning map $v_{\cdots\to\cx{\alpha}}$ admits a context changing map and an observable $\hat{O}$ is stable in the contexts $\cx{\alpha}$ and $\cx{\beta}$,
then
\begin{equation}
v_{\cdots\to\cx{\alpha}}(\hat{O}) (\canonical{\alpha}{\beta}^{-1}(\omega)) = v_{\cdots\to\cx{\alpha}\to\cx{\beta}}(\hat{O})(\omega),\qquad \mbox{ a.s. on } \PhaseSp.
\end{equation}
\end{proposition}

\Proof
By Proposition \ref{INVARIANT},
for all possible $\omega\in \PhaseSp$,
\begin{eqnarray*}
&& v_{\cdots\to\cx{\alpha}}(\hat{O}) (\canonical{\alpha}{\beta}^{-1}(\omega))\\
&\stackrel{(\ref{CONTCHG})}{=}& v_{\cdots\to\cx{\alpha}\to\cx{\beta}}(\Unitary{\alpha}{\beta}\ \hat{O}\ \uNitary{\alpha}{\beta})(\omega)\\
&=& v_{\cdots\to\cx{\alpha}\to\cx{\beta}}(\hat{O})(\omega).
\end{eqnarray*}
\QED

\begin{proposition}\label{FUNCproof}
Suppose that a value assigning map $v_{\cdots\to\cx{\alpha}}$ admits a context changing map $\canonical{\alpha}{\beta}$ from $\cx{\alpha}$ to $\cx{\beta}$.
Let $f$ be a function such that $\hat{A} = f(\hat{B})$ for some observables $\hat{A}$, $\hat{B}$  $\incx \cx{\alpha}$.
If $\hat{A} \incx \cx{\beta}$ and
\begin{equation}
v_{\cdots\to\cx{\alpha}}(\hat{A}) = f\left( v_{\cdots\to\cx{\alpha}}(\hat{B})  \right), \mbox{ a.s. on } \PhaseSp   \label{FUNCproof1}
\end{equation}
holds,
then
\begin{eqnarray}
\hat{A} &=& f(\Unitary{\alpha}{\beta} \hat{B} \uNitary{\alpha}{\beta}), \\
v_{\cdots\to\cx{\alpha}\to\cx{\beta}}(\hat{A}) &=& f\left( v_{\cdots\to\cx{\alpha}\to\cx{\beta}}(\Unitary{\alpha}{\beta} \hat{B} \uNitary{\alpha}{\beta})  \right), \mbox{ a.s. on } \PhaseSp.
\end{eqnarray}
\end{proposition}

\Proof
Since $\hat{A} \incx \cx{\alpha}$, $\cx{\beta}$, we have by Proposition \ref{INVARIANT}
\[
f\left( \Unitary{\alpha}{\beta} \hat{B} \uNitary{\alpha}{\beta} \right) 
=
\Unitary{\alpha}{\beta} \ \hat{A} \ \uNitary{\alpha}{\beta} = \hat{A}.
\]
By (\ref{CONTCHG}),
\begin{eqnarray*}
&&v_{\cdots\to\cx{\alpha}\to\cx{\beta}}(  \hat{A}  ) ( \omega ) \\
&=& v_{\cdots\to\cx{\alpha}\to\cx{\beta}}( \Unitary{\alpha}{\beta}\ \hat{A}\ \uNitary{\alpha}{\beta}  )(\omega) \\
&\stackrel{(\ref{CONTCHG})}{=}& v_{\cdots\to\cx{\alpha}}(  \hat{A}   )(\canonical{\alpha}{\beta}^{-1}(\omega)) \\
&\stackrel{(\ref{FUNCproof1})}{=}& f\left( v_{\cdots\to\cx{\alpha}}(\hat{B})(\canonical{\alpha}{\beta}^{-1}(\omega))  \right) \\
&\stackrel{(\ref{CONTCHG})}{=}& f\left( v_{\cdots\to\cx{\alpha}\to\cx{\beta}}( \Unitary{\alpha}{\beta}\ \hat{B}\ \uNitary{\alpha}{\beta})(\omega)  \right).
\end{eqnarray*}
\QED

\begin{proposition}\label{SisPERMUTATION}
For different contexts $\cx{\alpha} \equiv (|\alpha_{1}\>, \ldots, |\alpha_{n}\>)$, $\cx{\beta} \equiv (|\beta_{1}\>, \ldots, |\beta_{n}\>) $, $\cx{\gamma} \equiv (|\gamma_{1}\>, \ldots, |\gamma_{n}\>)$, define
\begin{equation}
S_{\cx{\alpha}}(\cx{\beta}\to\cx{\gamma}) := \uNitary{\alpha}{\gamma}\ \Unitary{\beta}{\gamma}\ \Unitary{\alpha}{\beta}.  \label{DefS}
\end{equation}
Then $S_{\cx{\alpha}}(\cx{\beta}\to\cx{\gamma})$ is a permutation of the CONS $(|\alpha_{1}\>, \ldots, |\alpha_{n}\>)$, i.e., there exists a permutation $p$ of the index set $\{1, \ldots, n\}$ such that
\begin{equation}
 |\alpha_{p(i)}\> = S_{\cx{\alpha}}(\cx{\beta}\to\cx{\gamma}) |\alpha_{i}\>,\ i=1, \ldots, n.
\end{equation}
\end{proposition}
\Proof
By definition of $\Unitary{\alpha}{\beta}$, for each $i$ there is $j$ such that $|\beta_{j}\>=\Unitary{\alpha}{\beta}|\alpha_{i}\>$.
Similarly, there exist $k$ and $l$ such that $|\gamma_{k}\>=\Unitary{\beta}{\gamma}|\beta_{j}\>$ and $|\alpha_{l}\>=\uNitary{\alpha}{\gamma}|\gamma_{k}\>$.
Hence $S_{\cx{\alpha}}(\cx{\beta}\to\cx{\gamma}) |\alpha_{i}\> =  |\alpha_{l}\>$, and therefore $p(i) = l$.
It is clear that this $p$ is a permutation of the index set.
\QED

\begin{proposition}\label{REDUCTION}
For a history of context change $\cx{\alpha}\to\cx{\xi}\to \cdots \to\cx{\zeta}\to \cx{\beta}$, there exists a permutation $p$ of the index set $\{1, \ldots, n\}$ such that
\begin{equation}
\Unitary{\zeta}{\beta} \cdots  \Unitary{\alpha}{\xi} |\alpha_{i}\> = \Unitary{\alpha}{\beta} |\alpha_{p(i)}\>,\ i= 1, \ldots, n.    \label{REDUCTIONofHISTORY}
\end{equation}
\end{proposition}
\Proof
In the history of context change $\cx{\alpha}\to\cx{\xi}\to\cx{\eta}\to \cdots \to\cx{\upsilon}\to\cx{\zeta}\to \cx{\beta}$, $|\alpha_{i}\>$ is mapped to
\[
\Unitary{\zeta}{\beta} \Unitary{\upsilon}{\zeta} \cdots \Unitary{\xi}{\eta} \Unitary{\alpha}{\xi} |\alpha_{i}\>.
\]
Using the notation $S_{\cx{\alpha}}(\cx{\xi}\to\cx{\eta})$  defined as (\ref{DefS}),
\begin{eqnarray*}
&&\Unitary{\zeta}{\beta} \Unitary{\upsilon}{\zeta} \cdots \Unitary{\xi}{\eta} \Unitary{\alpha}{\xi}\\
&=&\Unitary{\zeta}{\beta} \Unitary{\upsilon}{\zeta} \cdots \Unitary{\alpha}{\eta} S_{\cx{\alpha}}(\cx{\xi}\to\cx{\eta})\\
&=&\Unitary{\zeta}{\beta} \Unitary{\alpha}{\zeta} S_{\cx{\alpha}}(\cx{\upsilon}\to\cx{\zeta}) \cdots  S_{\cx{\alpha}}(\cx{\xi}\to\cx{\eta})\\
&=&\Unitary{\alpha}{\beta}  S_{\cx{\alpha}}(\cx{\zeta}\to\cx{\beta}) S_{\cx{\alpha}}(\cx{\upsilon}\to\cx{\zeta}) \cdots  S_{\cx{\alpha}}(\cx{\xi}\to\cx{\eta}).
\end{eqnarray*}
Thus we obtain
\begin{equation}
\uNitary{\alpha}{\beta} \Unitary{\zeta}{\beta} \Unitary{\upsilon}{\zeta} \cdots \Unitary{\xi}{\eta} \Unitary{\alpha}{\xi} =  S_{\cx{\alpha}}(\cx{\zeta}\to\cx{\beta}) S_{\cx{\alpha}}(\cx{\upsilon}\to\cx{\zeta}) \cdots  S_{\cx{\alpha}}(\cx{\xi}\to\cx{\eta}).
\end{equation}
By successive applications of Proposition \ref{SisPERMUTATION},
there exist a permutation $p$ of the index set such that
\[
S_{\cx{\alpha}}(\cx{\zeta}\to\cx{\beta}) S_{\cx{\alpha}}(\cx{\upsilon}\to\cx{\zeta}) \cdots  S_{\cx{\alpha}}(\cx{\xi}\to\cx{\eta}) |\alpha_{i}\> = |\alpha_{p(i)}\>
\]
and it is clear that this $p$ satisfies (\ref{REDUCTIONofHISTORY}).
\QED

\begin{corollary}\label{BackToAlpha}
For a history of context change $\cx{\alpha}\to\cx{\xi}\to \cdots \to\cx{\zeta}\to \cx{\beta}$, there exists a permutation $p$ of the index set $\{1, \ldots, n\}$ such that
\begin{equation}
\uNitary{\alpha}{\xi} \cdots \uNitary{\zeta}{\beta}  |\beta_{i}\> = |\alpha_{p(i)}\>,\ i= 1, \ldots, n.
\end{equation}
\end{corollary}
\Proof
By Proposition \ref{REDUCTION}, there is a permutation $q$ of the index set such that
\begin{equation}
\Unitary{\zeta}{\beta} \cdots  \Unitary{\alpha}{\xi} |\alpha_{i}\> = \Unitary{\alpha}{\beta} |\alpha_{q(i)}\>,\ i= 1, \ldots, n.
\end{equation}
By definition of $\Unitary{\alpha}{\beta}$, there is a permutation $r$ of the index set such that $\Unitary{\alpha}{\beta}|\alpha_{j}\> = |\beta_{r(j)}\>$.
It suffices to put $p = q^{-1} \circ r^{-1}$.
\QED

\section{A mathematical realization in the case of a finite-dimensional Hilbert space}
It is necessary to show that there exists an example  that realized the idea presented in the previous section at least mathematically.
We shall show the following theorem.
\begin{theorem}
Let $\Hlbt$ be an $n$-dimensional complex Hilbert space ($2 \leq n < \infty$).
Let $|\varphi\> \in \Hlbt$ be an arbitrary quantum state vector.
Let $\Cntx$ be a set of contexts in $\Hlbt$.
Then for a sufficiently small positive real number $\epsilon > 0$, there are a symplectic manifold $\PhaseSp$ and a subset $B(\epsilon) \subset \PhaseSp$ such that
for an arbitrary history of context change $\cx{\alpha} \to \cdots \to \cx{\beta}$  that begins with a fixed context $\cx{\alpha}$,  there exist
\begin{enumerate}
\def\labelenumi{\roman{enumi})}
\item a one-to-one mapping  $\hSplit{ \cx{\alpha}\to\cdots\to\cx{\beta} }$
of $B(\epsilon)$ into $\PhaseSp$, \item a  probability
distribution $ P^{|\varphi\>;
\epsilon}_{\cx{\alpha}\to\cdots\to\cx{\beta}} $ on $\PhaseSp$ that
is obtained  under the mapping $\hSplit{
\cx{\alpha}\to\cdots\to\cx{\beta} }$  from a probability
distribution $\mu(|\varphi\>, \epsilon)$ that is uniform on
$B(\epsilon)$ and vanishes outside of $B(\epsilon)$;
hence
\begin{equation}
d \mu(|\varphi\>; \epsilon) = d P^{|\varphi\>; \epsilon}_{\cx{\alpha}\to\cdots\to\cx{\beta}} \circ \hSplit{ \cx{\alpha}\to\cdots\to\cx{\beta} },
\end{equation}
\item a value assigning map
$v_{\cx{\alpha}\to\cdots\to\cx{\beta}}$ such that the expectation
value of $v_{\cx{\alpha}\to\cdots\to\cx{\beta}}(\hat{O})$ for any
observable $\hat{O} \incx\cx{\beta}$ calculated with $
P^{|\varphi\>; \epsilon}_{\cx{\alpha}\to\cdots\to\cx{\beta}} $
coincides with the corresponding one provided by
quantum probability theory. And the random variable
$v_{\cx{\alpha}\to\cdots\to\cx{\beta}}(\hat{O})$ assumes only
eigenvalues  of $\hat{O}$ almost surely on $\Omega$.
\end{enumerate}
Moreover, these value assigning maps satisfy gFUNC and n-TRNS.
\end{theorem}

The proof consists of six steps.
In the first, we construct a symplectic manifold $\PhaseSp$ from $\Hlbt$ and define a volume element.
In the second, we define a probability distribution $\mu(|\varphi\>; \epsilon)$ on $\PhaseSp$ that represents an ensemble prepared before measurements.
In the third, we introduce a mapping called a {\it splitting} which maps prepared ensemble to the one just before measurements, and define  $P^{|\varphi\>; \epsilon}_{\cx{\alpha}\to\cdots\to\cx{\beta}}$ as the results of the splitting.
In the fourth, we define value assignments and show that gFUNC is satisfied.
In the fifth, we show that quantum-mechanical expectation values are reproduced.
In the final step, we show that n-TRNS is satisfied.\\

\Step{1}
Let us fix a complete orthonormal system $(\ |\alpha_{1}\>, \ldots, |\alpha_{n}\>\ )$ of $\Hlbt$ that specifies a context $\cx{\alpha} \in \Cntx$.
We define functions $x^{i}$, $y^{i}$ $(i = 1, 2, \ldots, n)$ of $\Hlbt$ into $\Real$ as the real part and the imaginary part of the coefficient of $|\alpha_{i}\>$ in expansion of a vector of $\Hlbt$ respectively, for $i = 1, 2, \ldots, n$.
Then for an arbitrary vector $|\varphi\> \in \Hlbt$,
\begin{eqnarray}
|\varphi\> = \sum_{i = 1}^{n}  \Bigl( x^{i}({|\varphi\>}) + \sqrt{-1} y^{i}({|\varphi\>})  \Bigr) |\alpha_{i}\>.
\end{eqnarray}
Then $( x^{1}, \ldots, x^{n},\ y^{1}, \ldots, y^{n})$ becomes a coordinate system of $\Hlbt$.

Let us define  $\PhaseSp = \Real^{2n}$.
We denote this coordinate system by $\coord$.
Then we define $\coord$ as
\begin{equation}
\coord : |\varphi\> \in \Hlbt \mapsto \Bigl( x^{1}(|\varphi\>), \ldots, x^{n}(|\varphi\>),\ y^{1}(|\varphi\>), \ldots, y^{n}(|\varphi\>) \Bigr) \in \PhaseSp.
\end{equation}

Using this coordinate system, we define a symplectic structure on $\Hlbt$ by a symplectic form $\sigma$ defined by
\begin{equation}
\sigma := \sum_{i = 1}^{n} d x^{i} \wedge d y^{i},   \label{SIMPLECTIC}
\end{equation}
where $\wedge$ means the exterior product.
For the sake of convenience, we use matrix notation:
${\bf x} := [x^{1} , \ldots, x^{n}]^T$, ${\bf y} := [y^{1} , \ldots, y^{n}]^T$, where $T$ means transpose.
Then the symplectic form can be written as
\begin{equation}
\sigma =  d {\bf x}^{T} \wedge d{\bf y}.
\end{equation}

\begin{lemma}\label{UNITARYisCANONICAL}
Let $U$ be a unitary transformation of an $n$-dimensional Hilbert space $\Hlbt$.
Then $U$ is a canonical transformation of $\Hlbt$ with respect to the symplectic structure $\sigma$ defined by (\ref{SIMPLECTIC}).
\end{lemma}

\Proof
Put $z^{i} := x^{i} + \sqrt{-1} y^{i}$ ($i = 1, \ldots, n$).
For convenience, let us use matrix notation;
${\bf z} := [z^{1} , \ldots, z^{n}]^T$, where $T$ represents transpose.
Since
\begin{eqnarray*}
\overline{z^{i}} d z^{i} &=& \left( x^{i} - \sqrt{-1} y^{i} \right) \left(d x^{i} + \sqrt{-1} d y^{i}  \right)\\
&=& \frac{1}{2} d (x^{i})^{2} + \frac{1}{2} d (y^{i})^{2} - \sqrt{-1}d (x^{i} y^{i}) + 2 \sqrt{-1} x^{i} d y^{i} ,
\end{eqnarray*}
By taking the exterior derivatives, we obtain
\begin{equation}
d ({\bf z}^{\dagger} d {\bf z}) = 2\sqrt{-1} d {\bf x}^{T} \wedge d {\bf y},  \label{COMPLEX}
\end{equation}
where $\dagger$ represents complex conjugate transpose.

We denote by ${\bf U}$  the $n\times n$-matrix whose $(i,j)$-entry is $ \<\alpha_{i}|U|\alpha_{j}\>$.
It is easy to see that for an arbitrary $|\varphi\>\in \Hlbt$
\begin{equation}
{\bf z}( U |\varphi\> ) = [z^{i}( U |\varphi\> )] = [\<\alpha_{i}|U|\varphi\> ]  = \left[ \sum_{j=1}^{n} \<\alpha_{i}|U|\alpha_{j}\> \<\alpha_{j} |\varphi\> \right] = {\bf U} {\bf z}(|\varphi\>).
\end{equation}
The pull-back of ${\bf z}^{\dagger} d {\bf z}$ under $U$ becomes
\begin{eqnarray*}
U^{*} ( {\bf z}^{\dagger} d {\bf z} ) &=& ({\bf U z})^{\dagger} d({\bf U z}) \\
&=& {\bf z}^{\dagger}{\bf U}^{\dagger}{\bf U} d {\bf z}\\
&=& {\bf z}^{\dagger} d {\bf z}.
\end{eqnarray*}
Thus ${\bf z}^{\dagger} d {\bf z}$ is invariant under $U$.
Since the symplectic form is equal to the exterior derivative of ${\bf z}^{\dagger} d {\bf z}$ except a scalar multiple as we saw in (\ref{COMPLEX}), $U$ is a canonical transformation of $\Hlbt$.
\QED

We define a volume element $d^{n} x d^{n} y$ of $\Hlbt$ by
\begin{equation}
d^{n}x d^{n}y := dx^{1} \wedge \ldots \wedge dx^{n} \wedge dy^{1} \wedge \ldots \wedge dy^{n} = {(-1)^{n(n-1)/2}}\frac{1}{n!} \wedge_{i= 1}^{n }\sigma.
\end{equation}

\begin{corollary}\label{VOLUME}
Let $U$ be a unitary transformation of an $n$-dimensional Hilbert space $\Hlbt$.
$U$ preserves the volume element $d^n x d^n y$.
\end{corollary}

\Proof
By Lemma \ref{UNITARYisCANONICAL}, $U$ preserves the symplectic form $\sigma$.
Hence the volume element $d^n x d^n y$ which is an $n$th power of $\sigma$ with respect to exterior product is preserved under $U$.
\QED

\Step{2}
We define an $\epsilon$-ball with a center $|\varphi\>\in \Hlbt$ by
\begin{equation}
B^{2n}( |\varphi\>; \epsilon) := \left\{ |f\> \in \Hlbt :\  \||\ |f\> - |\varphi\> \ \|| < \epsilon  \right\}.
\end{equation}
Let us consider an ensemble of vectors that are distributed uniformly on $B^{2n}(|\varphi\>; \epsilon )$.
Define $B(\epsilon) := \coord\left( B^{2n}(|\varphi\>; \epsilon ) \right)$.
By using $\coord$, this ensemble is characterized by a probability distribution $\mu(|\varphi\>; \epsilon )$ on $\PhaseSp$.
Using matrix notation of the coordinates $({\bf x}, {\bf y})$, we can write it as
\begin{equation}
d\mu(|\varphi\>; {\epsilon})(({\bf x},\ {\bf y})) :=
\frac{1}{ |B^{2n}(|\varphi\>; \epsilon)|} \chi_{B(\epsilon)}(({\bf x},\ {\bf y})) d^n x d^n y,
\end{equation}
where $\chi_{B(\epsilon)}$ is the characteristic function of $B(\epsilon)$ and $|B^{2n}( |\varphi\>; {\epsilon})|$ denotes the volume of $B^{2n}( |\varphi\>; {\epsilon})$.

\Step{3}
We call an injection $\split{\beta}$ of $B^{2n}(|\varphi\>; \epsilon)$ into  $\Hlbt$ a {\it splitting} in a context $\cx{\beta}$, if there exists a split $\{ D^{\epsilon}_{i}: i= 1, 2, \ldots , n  \}$ of $B^{2n}(|\varphi\>; \epsilon)$, i.e.,
 $B^{2n}(|\varphi\>; \epsilon) = \cup_{i=1}^{n}D^{\epsilon}_{i}$ and  $D_{i}^{\epsilon} \cap D_{j}^{\epsilon} = \emptyset $ for $i \not= j$,
and the volume of $D^{\epsilon}_{i}$ is proportional to $|\< \beta_{i}| \varphi\>|^2$, and
\begin{equation}
\left| \split{\beta}( D^{\epsilon}_{i} ) \ \Delta \  B^{2n}(|\beta_{i}\>; \epsilon |\<\beta_{i}|\varphi\>|^{1/n} ) \right| = 0
\end{equation}
($i = 1, 2, \ldots, n$).
Here, $X \ \Delta \ Y$ represents the symmetric difference of subsets $X$ and $Y$ defined by $X \ \Delta \ Y := (X - Y) \cup (Y- X)$.\\

\begin{lemma}\label{SPLITTING}
If $\epsilon > 0$ is sufficiently small, then there exists a splitting of $B^{2n}(|\varphi\>; \epsilon)$ in an arbitrary context $\cx{\beta} \in \Cntx.$
\end{lemma}

\Proof A split $\{ D^{\epsilon}_{i} \}$ of $B^{2n}(|\varphi\>;
\epsilon)$ is obtained by the following procedure. Let ${\bf
R}_{\coord(|\varphi\>)}(\theta)$ be a rotation around the
direction $\coord(|\varphi\>)$ with an angle $\theta$ ($0\leq
\theta < 2\pi$) in $\PhaseSp$ such that ${\bf
R}_{\coord(|\varphi\>)}(2\pi)$ = the identity mapping. Put
$g_{\theta} := \coord^{-1} \circ {\bf
R}_{\coord(|\varphi\>)}(\theta) \circ \coord$.

Let $B_{0}$ be a subset of $B^{2n}(|\varphi\>; \epsilon)$ such that
\begin{eqnarray*}
g_{\theta}(B_{0}) \cap g_{\theta'}(B_{0}) &=& \emptyset \ \mbox{ if } \theta \not= \theta', 0 \leq \theta, \theta' < 2\pi,\\
B^{2n}(|\varphi\>; \epsilon ) &=& \bigcup \left\{ g_{\theta}(B_{0}) :\ 0 \leq \theta < 2\pi \right\}.
\end{eqnarray*}
Let $\theta_{0} (= 0) < \theta_{1} < \ldots < \theta_{n} (= 2\pi)$
be an increasing sequence of angles around the direction
$\coord(\cket{\varphi})$ such that the volume of  $\{ g_{\theta}(
B_{0} ) :  \theta_{i-1} \leq \theta < \theta_{i}  \}$  is
proportional to  $|\<\beta_{i} | \varphi\>|^2$ ($i = 1, 2, \ldots,
n$). Put $D^{\epsilon}_{i} := \{ g_{\theta}( B_{0} ) :
\theta_{i-1} \leq \theta < \theta_{i}  \}$ ($i = 1, 2, \ldots,
n$). For sufficient small $\epsilon$, $B^{2n}(|\beta_{i}\>;
\epsilon|\<\beta_{i}|\varphi\>|^{1/n})$s are pairwise disjoint.
Then we can define a one-to-one mapping $\split{\beta}$ of
$B^{2n}(|\varphi\>; \epsilon)$ into $\Hlbt$ by defining an action
of $\split{\beta}$ on $D^{\epsilon}_{i}$ by the successive
compositions of a stretching $D^{\epsilon}_{i}$ along
$g_{\theta}$-direction so as to make it a $2n$-dimensional ball, a
contraction along radial direction of the ball, and a parallel
translation mapping the center of the ball from $|\varphi\>$ to
$|\beta_{i}\>$ ($i=1, \ldots, n$). This completes the proof. \QED
\\

First, we define  $P^{|\varphi\>}_{\cx{\alpha}}$ as the
probability distribution characterizing an ensemble that is
obtained by an action of the splitting $\split{\alpha}$ in
$\cx{\alpha}$ (Lemma \ref{SPLITTING}) on the ensemble
characterized by $\mu(|\varphi\>; \epsilon)$. Since
\begin{equation}
d \mu(|\varphi\>; \epsilon)(({\bf x}, {\bf y})) = dP^{|\varphi\>}_{\cx{\alpha}} \circ \coord \circ \split{\alpha} \circ \coord^{-1} (({\bf x}, {\bf y})),  \qquad \forall ({\bf x}, {\bf y}) \in \Real^{2n} = \Omega = \coord( \Hlbt ),
\end{equation}
\begin{equation}
d P^{|\varphi\>}_{\cx{\alpha}}(({\bf x},{\bf y})) = \frac{1}{|B^{2n}(|\varphi\>; \epsilon) |} \sum_{i = 1}^{n} \chi_{ \coord\left( B^{2n}(|\alpha_{i}\>; \epsilon |\<\alpha_{i}|\varphi\>|^{1/n} ) \right) }(({\bf x},{\bf y})) d^n x d^n y
\end{equation}
holds.

Suppose that up to $k$th context $\cx{\delta}$ of the history of context change ${\cx{\alpha}\to \cdots \to\cx{\delta}}$, a splitting $\hsplit{\cx{\alpha}\to \cdots \to\cx{\delta}}$ in a context $\cx{\delta}$ is defined;
$P^{|\varphi\>}_{\cx{\alpha}\to \cdots \to\cx{\delta}}$ is defined from $\mu(|\varphi\>; \epsilon)$ through $\hsplit{\cx{\alpha}\to \cdots \to\cx{\delta}}$.
Now our aim is to define $\hsplit{\cx{\alpha}\to \cdots \to\cx{\delta}\to\cx{\gamma}}$ in the $(k+1)$th context $\cx{\gamma}$ of the history of context change $\cx{\alpha}\to\cdots\to\cx{\beta}$.

We denote the split of $B^{2n}(|\varphi\>; \epsilon)$ with respect
to the splitting $\hsplit{\cx{\alpha}\to\cdots\to\cx{\delta}}$ by
$\{ D^{\epsilon}_{i}(\cx{\alpha}\to\cdots\to\cx{\delta}) :\ i = 1,
2, \ldots, n \}$. These $
D^{\epsilon}_{i}(\cx{\alpha}\to\cdots\to\cx{\delta})$s may be
different from the $D^{\epsilon}_{i}$s defined in the proof of
Lemma \ref{SPLITTING}, but by the assumption they satisfies the
following:
\begin{eqnarray}
\left| D^{\epsilon}_{i}(\cx{\alpha}\to\cdots\to\cx{\delta}) \right| &\propto& |\<\delta_{i}|\varphi\>|^2,\\
D^{\epsilon}_{i}(\cx{\alpha}\to\cdots\to\cx{\delta}) \cap D^{\epsilon}_{j}(\cx{\alpha}\to\cdots\to\cx{\delta}) &=& \emptyset \mbox{ if } i\not= j,\\
\bigcup_{i = 1}^{n} D^{\epsilon}_{i}(\cx{\alpha}\to\cdots\to\cx{\delta}) &=& B^{2n}(|\varphi\>; \epsilon),\\
\left| \hsplit{\cx{\alpha}\to\cdots\to\cx{\delta}} \left( D^{\epsilon}_{i}(\cx{\alpha}\to\cdots\to\cx{\delta}) \right) \right.  &\Delta&  \left. B^{2n}\left( |\delta_{i}\>;\  \epsilon |\<\delta_{i}|\varphi\>|^{1/n}\right) \right| = 0.
\end{eqnarray}
By Lemma \ref{FINEST}, there exist the finest partitions $\{ I_{i} \}$ and $\{ J_{j}\}$ of the index set $\{1, \ldots, n\}$ such that $\span \{ |\delta_{i} \> :\ i \in I_{k} \} = \span \{ |\gamma_{j}\> :\ j \in J_{k} \}$.
By splitting $D^{\epsilon}_{i}(\cx{\alpha}\to\cdots\to\cx{\delta})$s finer if necessary and collecting them, we can define a split $\{ D^{\epsilon}_{i}(\cx{\alpha}\to\cdots\to\cx{\delta}\to\cx{\gamma})  \}$ of $B^{2n}(|\varphi\>; \epsilon)$ so that
\begin{eqnarray}
\left| \bigcup_{j\in J_{k} } D^{\epsilon}_{j}(\cx{\alpha}\to\cdots\to\cx{\delta}\to\cx{\gamma}) \right. &\Delta& \left. \bigcup_{i \in I_{k} } D^{\epsilon}_{i}(\cx{\alpha}\to\cdots\to\cx{\delta}) \right| = 0,
\label{NewD} \\
\left| D^{\epsilon}_{i}(\cx{\alpha}\to\cdots\to\cx{\delta}\to\cx{\gamma}) \right| &\propto& |\<\gamma_{i}|\varphi\>|^2,\\
D^{\epsilon}_{i}(\cx{\alpha}\to\cdots\to\cx{\delta}\to\cx{\gamma}) \cap D^{\epsilon}_{j}(\cx{\alpha}\to\cdots\to\cx{\delta}) &=& \emptyset \mbox{ if } i\not= j,\\
\bigcup_{i = 1}^{n} D^{\epsilon}_{i}(\cx{\alpha}\to\cdots\to\cx{\delta}\to\cx{\gamma}) &=& B^{2n}(|\varphi\>; \epsilon),\\
\left| \hsplit{\cx{\alpha}\to\cdots\to\cx{\delta}\to\cx{\gamma}} \left( D^{\epsilon}_{i}(\cx{\alpha}\to\cdots\to\cx{\delta}\to\cx{\gamma}) \right) \right. & \Delta & \left.  B^{2n}\left( |\gamma_{i}\>;\  \epsilon |\<\gamma_{i}|\varphi\>|^{1/n}\right) \right| = 0
\end{eqnarray}
hold.
We define the splitting $\hsplit{\cx{\alpha}\to\cdots\to\cx{\delta}\to\cx{\beta}}$ as an injection of $B^{2n}(|\varphi\>; \epsilon)$ into $\Hlbt$ that satisfies
\begin{equation}
\hsplit{\cx{\alpha}\to\cdots\to\cx{\delta}\to\cx{\beta}}\left( D^{\epsilon}_{i}(\cx{\alpha}\to\cdots\to\cx{\delta}\to\cx{\beta}) \right)
=  B^{2n}\left( |\beta_{i}\>;\  \epsilon |\<\beta_{i}|\varphi\>|^{1/n}\right)
\label{DefSplit}
\end{equation}
for $i = 1, \ldots, n$.

We define  $P^{|\varphi\>}_{\cx{\alpha}\to \cdots \to \cx{\gamma}}$ as the probability distribution characterizing the  ensemble that is obtained as results of action of the splitting $\hsplit{\cx{\alpha}\to\cdots\to\cx{\gamma}}$ on the ensemble characterized by $\mu(|\varphi\>; \epsilon)$.
Since
\begin{equation}
d \mu(|\varphi\>; \epsilon)(({\bf x}, {\bf y})) = dP^{|\varphi\>}_{\cx{\alpha}\to \cdots \to\cx{\gamma}} \circ  \coord \circ \hsplit{\cx{\alpha}\to\cdots\to\cx{\gamma}} \circ \coord^{-1} (({\bf x}, {\bf y})),  \qquad \forall ({\bf x}, {\bf y}) \in \Real^{2n} = \Omega =  \coord (\Hlbt),
\end{equation}
\begin{equation}
d P^{|\varphi\>}_{\cx{\alpha}\to \cdots \to\cx{\gamma}}(({\bf x},{\bf y})) = \frac{1}{|B^{2n}(|\varphi\>; \epsilon) |} \sum_{i = 1}^{n} \chi_{ \coord \left( B^{2n}(|\gamma_{i}\>; \epsilon |\<\gamma_{i}|\varphi\>|^{1/n} ) \right) }\left(({\bf x},{\bf y})\right) d^n x d^n y.
\label{DefP}
\end{equation}
holds.
It suffices to put $ \hSplit{\cx{\alpha}\to\cdots\to\cx{\gamma}} := \coord \circ \hsplit{\cx{\alpha}\to\cdots\to\cx{\gamma}} \circ \coord^{-1}$.
By repeating these constructions  from $k=1$ we obtain $\hsplit{\cx{\alpha}\to\cdots\to\cx{\beta}}$, $P^{|\varphi\>}_{\cx{\alpha}\to \cdots \to \cx{\beta}}$, and  $\hSplit{\cx{\alpha}\to\cdots\to\cx{\beta}}$.\\

\Step{4}
For each pair of different contexts $\cx{\delta}$, $\cx{\gamma} \in \Cntx$, as the counterpart of the unitary transformation $\Unitary{\delta}{\gamma}$ of context change, we define $\canonical{\delta}{\gamma} : \PhaseSp\to\PhaseSp$  by
\begin{equation}
\canonical{\delta}{\gamma} := \coord \circ \Unitary{\delta}{\gamma} \circ \coord^{-1}.
\end{equation}

First, we define a value assigning map $v_{\cx{\alpha}}( \cdot )$ in the context $\cx{\alpha}$ that corresponds to the orthonormal basis $\{ |\alpha_{1}\>, \ldots, |\alpha_{n}\>  \}$ in the following way.
Let $\hat{O}$ be an observable that is stable in  $\cx{\alpha}$.
We denote the eigenvalue of $\hat{O}$ associated with an eigenvector $| \alpha_{i}\>$ by $o_{i}$.
Let $\{I'_{1}, \ldots, I'_{m} \}$ be a partition of the index set $\{1, 2, \ldots, n\}$ such that $o_{i} = o_{j}$ iff $i, j \in I'_{k}$ for some $k$.
For the sake of convenience, we write $o_{I'_{k}}$ instead of $o_{i}$ whose index $i$ belongs to $I'_{k}$.
We define $v_{\cx{\alpha}}$ by
\begin{equation}
v_{\cx{\alpha}} ( \hat{O} )( ({\bf x}, {\bf y}) ) := \left\{
\begin{array}{l l}
o_{I'_{k}}, & \mbox{ if } \coord^{-1}({\bf x}, {\bf y}) \in \bigcup \left\{ B^{2n}(|f\>; \epsilon)  :\ \||\ |f\>\ \|| = 1,\  |f\> \in \span\left\{ |\alpha_{i}\> :\ i \in I'_{k} \right\} \right\}, \\
\ast,  & \mbox{ otherwise, }
\end{array}
\right.
\end{equation}
where $\ast$ represents a value depending on $({\bf x}, {\bf y})$.

For each context $\cx{\delta} \in \Cntx$, we define $v_{\cx{\alpha}\to\cx{\delta}}$ for $\hat{O}\incx\cx{\delta}$ by
\begin{equation}
v_{\cx{\alpha}\to\cx{\delta}}( \hat{O})(({\bf x}, {\bf y})) := v_{\cx{\alpha}}( \uNitary{\alpha}{\delta}\ \hat{O}\ \Unitary{\alpha}{\delta}  )( \canonical{\alpha}{\delta}^{-1}(({\bf x}, {\bf y})) ),  \quad \forall ({\bf x}, {\bf y}) \in \Real^{2n} = \Omega = \coord( \Hlbt ).
\end{equation}

In more general case, we define value assigning maps successively.
Suppose that up to $k$th context $\cx{\delta}$ of the history of context change ${\cx{\alpha}\to \cdots \to\cx{\delta}}$, $v_{\cx{\alpha}\to\cdots\to\cx{\delta}}$ is defined.
For the $(k+1)$th context $\cx{\gamma}$, we define $v_{\cx{\alpha}\to\cdots\to\cx{\delta}\to\cx{\gamma}}$
for $\hat{O} \incx\cx{\gamma}$ by
\begin{equation}
v_{\cx{\alpha}\to\cdots\to\cx{\delta}\to\cx{\gamma}}( \hat{O} )(({\bf x}, {\bf y})) := v_{\cx{\alpha}\cdots\to\cx{\delta}}(\uNitary{\delta}{\gamma}\ \hat{O}\ \Unitary{\delta}{\gamma})( \canonical{\delta}{\gamma}^{-1}(({\bf x}, {\bf y})) ),  \quad \forall ({\bf x}, {\bf y}) \in \Real^{2n} = \Omega = \coord( \Hlbt ).
\end{equation}

By definition, these value assigning maps admit context changing maps.

\begin{proposition}\label{BALL}
Let $U$ be a unitary transformation of $\Hlbt$.
\begin{equation}
U B^{2n}( |\varphi\> ; r ) = B^{2n}( U |\varphi\>; r),
\end{equation}
where $|\varphi\> \in \Hlbt$, $r > 0$.
\end{proposition}

\Proof
\begin{eqnarray*}
&&U B^{2n}( |\varphi\> ; r )\\
&=& U \left\{ |f\> \in \Hlbt :\  \||\ |f\> - |\varphi\> \ \|| < r \right\}\\
&=& \left\{ U|f\> \in \Hlbt :\  \||\ |f\> - |\varphi\> \ \|| < r \right\}\\
&=& \left\{ |g\> \in \Hlbt :\  \||\ U^{\dagger}|g\> - |\varphi\> \ \|| < r \right\}\\
&=& \left\{ |g\> \in \Hlbt :\  \||\ |g\> - U|\varphi\> \ \|| < r \right\}\\
&=& B^{2n}( U|\varphi\> ; r ).
\end{eqnarray*}
\QED

\begin{proposition}
The value assigning maps $v_{\cx{\alpha}\to\cdots\to\cx{\beta}}( \hat{O} ) \circ \hSplit{\cx{\alpha}\to\cdots\to\cx{\beta}}$s defined in the above satisfy gFUNC.
\end{proposition}
\Proof Let $\hat{A}$ and $\hat{B}$ be observables such that there
exists a function $f: \Real \to \Real$ which satisfies $\hat{A} =
f(\hat{B})$. Suppose that $\hat{B}$ is stable in a context
$\cx{\beta} \in \Cntx$. For a history of context changes
$\cx{\alpha}\to \cx{\xi} \to \cdots \to \cx{\zeta} \to
\cx{\beta}$, there exist unitary transformations
$\Unitary{\alpha}{\xi}, \ldots, \Unitary{\zeta}{\beta}$ of context
changes corresponding to each step. It is easy to see that
\begin{eqnarray*}
\uNitary{\alpha}{\xi} \cdots \uNitary{\zeta}{\beta}
\hat{B}
\Unitary{\zeta}{\beta} \cdots \Unitary{\alpha}{\xi} =: \hat{O} &\incx& \cx{\alpha},\\
\uNitary{\alpha}{\xi} \cdots \uNitary{\zeta}{\beta}
\hat{A}
\Unitary{\zeta}{\beta} \cdots \Unitary{\alpha}{\xi} =: \hat{Q} &\incx& \cx{\alpha},
\end{eqnarray*}
and
\[
\hat{Q}:= f(\hat{O}).
\]

Let $({\bf x}, {\bf y})$ be an arbitrary point in $B(\epsilon) = \coord( B^{2n}(|\varphi\>; \epsilon ))$.
There exists $i \in \{1, \ldots, n\}$  such that $\hSplit{\cx{\alpha}\to\cdots\to\cx{\beta}}(({\bf x}, {\bf y})) \in \coord( B^{2n}(|\beta_{i}\>; \epsilon|\<\beta_{i}|\varphi\>|^{1/n}) )$ in the notation in the step 3.
Put $|f\> := \coord^{-1}(({\bf x}, {\bf y}))$.
Let $a_{i}$ and $b_{i}$ be  eigenvalues of $\hat{A}$ and $\hat{B}$ associated with the eigenvector $|\beta_{i}\>$, respectively.
Then
\[
v_{\cx{\alpha}\to\cdots\to\cx{\beta}}(\hat{A})\circ \hSplit{\cx{\alpha}\to\cdots\to\cx{\beta}} (({\bf x}, {\bf y}))
= v_{\cx{\alpha}\to\cdots\to\cx{\beta}}(\hat{A})\circ \coord \circ \hsplit{\cx{\alpha}\to\cdots\to\cx{\beta}} (|f\>)
= a_{i}.
\]
In the same way,
\[
v_{\cx{\alpha}\to\cdots\to\cx{\beta}}(\hat{B})\circ \hSplit{\cx{\alpha}\to\cdots\to\cx{\beta}} (({\bf x}, {\bf y})) = b_{i}.
\]
Now our task is to show $a_{i} = f(b_{i})$.

By the definition of $v_{\cx{\alpha}\to\cdots\to\cx{\beta}}$ and (\ref{DefS}),
\begin{eqnarray*}
a_{i} &=& v_{\cx{\alpha}\to\cdots\to\cx{\beta}}(\hat{A}) \circ \hSplit{\cx{\alpha}\to\cdots\to\cx{\beta}} (({\bf x}, {\bf y}))\\
 &=& v_{\cx{\alpha}}( \uNitary{\alpha}{\xi} \cdots \uNitary{\zeta}{\beta} \hat{A} \Unitary{\zeta}{\beta} \cdots \Unitary{\alpha}{\xi}) \circ \canonical{\alpha}{\xi}^{-1} \circ \cdots \circ \canonical{\zeta}{\beta}^{-1} \circ
 \hSplit{\cx{\alpha}\to\cdots\to\cx{\beta}} (({\bf x}, {\bf y})) \\
&=& v_{\cx{\alpha}}( \hat{Q} ) \circ \canonical{\alpha}{\xi}^{-1} \circ \cdots \circ \canonical{\zeta}{\beta}^{-1} \circ \coord \circ \hsplit{\cx{\alpha}\to\cdots\to\cx{\beta}} (|f\>) \\
&=& v_{\cx{\alpha}}( \hat{Q} ) \circ \coord(  \uNitary{\alpha}{\xi}  \cdots \uNitary{\zeta}{\beta} \hsplit{\cx{\alpha}\to\cdots\to\cx{\beta}} (|f\>) ).
\end{eqnarray*}

By Corollary \ref{BackToAlpha} and Proposition \ref{BALL}, there
exists a permutation $p'$ of the index set $\{1, \ldots, n\}$ such
that $\uNitary{\alpha}{\xi}  \cdots \uNitary{\zeta}{\beta}
\hsplit{\cx{\alpha}\to\cdots\to\cx{\beta}} (|f\>)$ belongs to
$B^{2n}(|\alpha_{p'(i)}\>;
\epsilon|\<\beta_{i}|\varphi\>|^{1/n})$. Hence $a_{i}$ is an
eigenvalue of $\hat{Q}$ associated with an eigenvector
$|\alpha_{p'(i)}\>$. Since $\hat{Q} = f(\hat{O})$,
$|\alpha_{p'(i)}\>$ is an eigenvector of $\hat{O}$ with an
eigenvalue $o_{i}$ such that $a_{i} = f(o_{i})$. By definition of
$v_{\cx{\alpha}}$, the assigned value of $\hat{O}$ is $o_{i}$.
Therefore
\begin{eqnarray*}
a_{i}&=& v_{\cx{\alpha}}( f(\hat{O}) ) \circ \coord(
\uNitary{\alpha}{\xi}  \cdots \uNitary{\zeta}{\beta} \hsplit{\cx{\alpha}\to\cdots\to\cx{\beta}} (|f\>)
)\\
&=& f \circ  v_{\cx{\alpha}}( \hat{O} ) \circ \coord(
\uNitary{\alpha}{\xi}  \cdots \uNitary{\zeta}{\beta} \hsplit{\cx{\alpha}\to\cdots\to\cx{\beta}} (|f\>)
)\\
&=& f \circ v_{\cx{\alpha}}( \hat{O} )  \circ \canonical{\alpha}{\xi}^{-1} \circ \cdots \circ \canonical{\zeta}{\beta}^{-1} \circ
 \hSplit{\cx{\alpha}\to\cdots\to\cx{\beta}} (({\bf x}, {\bf y})) \\
&=& f \circ v_{\cx{\alpha}\to\cdots\to\cx{\beta}}(  \Unitary{\zeta}{\beta} \cdots \Unitary{\alpha}{\xi}  \hat{O} \uNitary{\alpha}{\xi} \cdots \uNitary{\zeta}{\beta} )  \circ  \hSplit{\cx{\alpha}\to\cdots\to\cx{\beta}} (({\bf x}, {\bf y})) \\
&=& f \circ v_{\cx{\alpha}\to\cdots\to\cx{\beta}}(   \hat{B}  )  \circ  \hSplit{\cx{\alpha}\to\cdots\to\cx{\beta}} (({\bf x}, {\bf y})) \\
&=& f(b_{i}).
\end{eqnarray*}
\QED

\Step{5}

\begin{proposition}
Let $\Borel$ be the Borel $\sigma$-algebra of $\PhaseSp$.
$(\PhaseSp, \Borel, P^{|\varphi\>; \epsilon}_{\cx{\alpha}\to \cdots \to \cx{\beta}})$ is a probability space that reproduce quantum-mechanical statistical results, i.e.,
for an observable $\hat{B} \incx \cx{\beta}$,
\begin{equation}
\int_{\PhaseSp} dP^{|\varphi\>; \epsilon}_{\cx{\alpha}\to \cdots \to\cx{\beta}} \ v_{\cx{\alpha}\to \cdots \to \cx{\beta}}(\hat{B}) = \<\varphi| \hat{B} | \varphi\>.
\end{equation}
\end{proposition}

\Proof
For the history of context change $\cx{\alpha} \to \cx{\gamma} \to \cdots \to \cx{\delta} \to \cx{\beta}$, there exists a sequence of unitary transformations of context changes $\Unitary{\alpha}{\gamma}, \ldots, \Unitary{\delta}{\beta}$ such that
$\uNitary{\alpha}{\gamma} \cdots \uNitary{\delta}{\beta} \hat{B} \Unitary{\delta}{\beta} \cdots \Unitary{\alpha}{\gamma} =: \hat{O}$ is stable in the context $\cx{\alpha}$.
Since  $\hat{B} = \sum_{i = 1}^{n} b_{i} |\beta_{i}\>\<\beta_{i}|$,  by using Corollary \ref{BackToAlpha}, we obtain
\begin{eqnarray*}
\hat{O} &=& \uNitary{\alpha}{\gamma}  \cdots \uNitary{\delta}{\beta}\  \hat{B}\ \Unitary{\delta}{\beta} \cdots \Unitary{\alpha}{\gamma} \\
&=& \sum_{i = 1}^{n} b_{i} \uNitary{\alpha}{\gamma}  \cdots \uNitary{\delta}{\beta}\  |\beta_{i}\>\<\beta_{i}|\ \Unitary{\delta}{\beta} \cdots \Unitary{\alpha}{\gamma} \\
&=&  \sum_{i = 1}^{n} b_{i}  |\alpha_{p(i)}\>\<\alpha_{p(i)}|,
\label{Ospe}
\end{eqnarray*}
where $p$ represents the permutation of the index set given by Corollary \ref{BackToAlpha}.

By the definition of $v_{\cx{\alpha}\to \cdots \to \cx{\beta}}(\hat{B})$,
\begin{equation}
v_{\cx{\alpha}\to \cdots \to \cx{\beta}}(\hat{B}) = v_{\cx{\alpha}}(\hat{O})\circ \canonical{\alpha}{\gamma}^{-1} \circ \cdots \circ \canonical{\delta}{\beta}^{-1}.
\end{equation}
Using Corollary \ref{BackToAlpha} again and (\ref{Ospe}), we have
\begin{eqnarray*}
v_{\cx{\alpha}\to \cdots \to \cx{\beta}}(\hat{B})(\coord( |\beta_{i}\>)) &=& v_{\cx{\alpha}}(\hat{O}) \left(  \coord(  \uNitary{\alpha}{\gamma}  \cdots  \uNitary{\delta}{\beta} |\beta_{i}\> ) \right)\\
&=& v_{\cx{\alpha}}(\hat{O}) \left(  \coord( |\alpha_{p(i)}\> ) \right)\\
&=& b_{i}.
\end{eqnarray*}
Therefore by (\ref{DefP})
\begin{eqnarray*}
&& \int_{\PhaseSp} d P^{|\varphi\>; \epsilon}_{\cx{\alpha} \to \cdots \to \cx{\beta}} \ v_{\cx{\alpha}\to \cdots \to \cx{\beta}}(\hat{B}) \\
 &=&  \frac{1}{|B^{2n}(|\varphi\>; \epsilon) |} \int_{\PhaseSp} d^n x d^n y \sum_{i = 1}^{n} \chi_{\coord \left( B^{2n}(|\beta_{i}\>; \epsilon |\<\beta_{i}|\varphi\>|^{1/n} ) \right)}
\ v_{\cx{\alpha}\to \cdots \to \cx{\beta}}(\hat{B})\\
 &=&   \sum_{i = 1}^{n} b_{i}
\frac{ | B^{2n}( |\beta_{i}\>; \epsilon |\<\beta_{i}|\varphi\>|^{1/n} )| }{|B^{2n}(|\varphi\>; \epsilon) |}\\
 &=&   \sum_{i = 1}^{n} b_{i} |\<\beta_{i}|\varphi\>|^{2}\\
 &=& \< \varphi| \hat{B} | \varphi\>.
\end{eqnarray*}
\QED
\\

\Step{6}
Finally, we show that n-TRNS is satisfied.

\begin{proposition}
For the history of context change $\cx{\alpha} \to \cdots \to \cx{\delta} \to \cx{\beta}$,
let $\hat{O}$ be a degenerate observable that is stable in both $\cx{\delta}$ and $\cx{\beta}$.
Then
\begin{equation}
v_{\cx{\alpha}\to \cdots \to \cx{\delta}\to\cx{\beta}}( \hat{O} ) \left(\coord(  \hsplit{\cx{\alpha}\to\cdots\to\cx{\delta}\to\cx{\beta}}( |f\> ) ) \right)
=  v_{\cx{\alpha}\to \cdots \to \cx{\delta}}(  \hat{O} ) \left(\coord(   \hsplit{\cx{\alpha}\to\cdots\to\cx{\delta}}( |f\> ) )\right)
\end{equation}
for $\forall |f\> \in B^{2n}(|\varphi\>; \epsilon)$.
\end{proposition}
\Proof
We use the same notations in the step 3.
If $|f\> \in D^{\epsilon}_{i}(\cx{\alpha}\to\cdots\to\cx{\delta}) \subset B^{2n}(\cket{\varphi}; \epsilon)$, then
\[
\hsplit{\cx{\alpha}\to\cdots\to\cx{\delta}}(|f\>) \in B^{2n}\left( |\delta_{i}\>; \epsilon|\<\delta_{i}|\varphi\>|^{1/n} \right).
\]
By the definition of the value assigning map $v_{\cx{\alpha}\to \cdots \to \cx{\delta}}$, $v_{\cx{\alpha}\to \cdots \to \cx{\delta}}(  \hat{O} )$ assumes an eigenvalue of $\hat{O}$, say $o_{k_{0}}$, i.e.,
\[
o_{k_{0}} =  v_{\cx{\alpha}\to \cdots \to \cx{\delta}}(  \hat{O} ) \left(\coord(   \hsplit{\cx{\alpha}\to\cdots\to\cx{\delta}}(|f\>) ) \right).
\]
Let $\{ I_{k} \}$ and $\{ J_{k} \}$ be the finest partitions of the index set with respect to the pair of contexts $\cx{\delta}$ and $\cx{\beta}$ (Lemma \ref{FINEST}).
Let $\Unitary{\delta}{\beta}$ be the unitary transformation of the context change from $\cx{\delta}$ to $\cx{\beta}$.
There exists $k$ such that $i\in I_{k}$.
For this $k$,  by (\ref{NewD}), there exists $j\in J_{k}$ such that $|\beta_{j}\>=\Unitary{\delta}{\beta}|\delta_{i}\>$ and $|f\> \in D^{\epsilon}_{j}(\cx{\alpha}\to\cdots\to\cx{\delta}\to\cx{\beta})$.
Hence by (\ref{DefSplit})
\[
\hsplit{\cx{\alpha}\to\cdots\to\cx{\delta}\to\cx{\beta}}(|f\>) \in B^{2n}\left( |\beta_{j}\>; \epsilon|\<\beta_{j}|\varphi\>|^{1/n} \right).
\]
By Proposition \ref{BALL},
\[
\uNitary{\delta}{\beta} \hsplit{\cx{\alpha}\to\cdots\to\cx{\delta}\to\cx{\beta}}(|f\>) \in B^{2n}\left( |\delta_{i}\>; \epsilon|\<\beta_{j}|\varphi\>|^{1/n} \right).
\]
By Proposition \ref{STABLE}, $v_{\cx{\alpha}\to \cdots \to \cx{\delta}\to\cx{\beta}}( \hat{O} )
=v_{\cx{\alpha}\to \cdots \to \cx{\delta}}( \hat{O} ) \circ \canonical{\delta}{\beta}^{-1}$, and therefore
\begin{eqnarray*}
&& v_{\cx{\alpha}\to \cdots \to \cx{\delta}\to\cx{\beta}}( \hat{O} ) \left(\coord(  \hsplit{\cx{\alpha}\to\cdots\to\cx{\delta}\to\cx{\beta}}(|f\>) ) \right) \\
&=& v_{\cx{\alpha}\to \cdots \to \cx{\delta}}( \hat{O} ) \left(\coord( \uNitary{\delta}{\beta} \hsplit{\cx{\alpha}\to\cdots\to\cx{\delta}\to\cx{\beta}}(|f\>) ) \right)\\
&=& o_{k_{0}}.
\end{eqnarray*}
\QED

Hence n-TRNS holds.
Thus the proof of the theorem is completed.

\section{Discussion}
Theorem 1 seemes to contradict to the no-go theorem for noncontextual hidden variable models \cite{KochenSpecker} by the following consideration.
By the no-go theorem, there exists a finite set of contexts $\{ \cx{\alpha(j)} : j = 1, \ldots, N\}$ such that it is impossible to define a value assigning map $v$ for them, if the dimension of the Hilbert space of quantum state vectors is greater than two.
For each context $\cx{\alpha(j)}$, there exists nondegenerate observable $\hat{A}(j) \incx \cx{\alpha(j)}$.
Let $\hat{D}$ be an observable that cannot be assigned a value.

Suppose that $\hat{D}$ is stable in every contexts $\cx{\alpha(1)}, \ldots, \cx{\alpha(N)}$.
Then there exists a function $f^{j}$ such that $\hat{D} = f^{j}(\hat{A}(j))$.
By gFUNC, 
\[
f^{k}\left(v_{\cx{\alpha(1)} \to \cdots \to\cx{\alpha(k)}}(\hat{A}(k)) \right) 
= v_{\cx{\alpha(1)} \to \cdots \to\cx{\alpha(k)}}(\hat{D})
\]
for an arbitrary history of context change $\cx{\alpha(1)} \to \cdots \to\cx{\alpha(k)}$ ($k=1, \ldots, N$).
By n-TRN,
\[
v_{\cx{\alpha(1)}}(\hat{D})
=
\cdots
=
v_{\cx{\alpha(1)} \to \cdots \to\cx{\alpha(k)}}(\hat{D})
=
\cdots
=
v_{\cx{\alpha(1)} \to \cdots \to\cx{\alpha(N)}}(\hat{D}),
\]
and
therefore
\[
f^{k}\left(v_{\cx{\alpha(1)} \to \cdots \to\cx{\alpha(k)}}(\hat{A}(k)) \right) 
= v_{\cx{\alpha(1)} \to \cdots \to\cx{\alpha(k)}}(\hat{D})
= v_{\cx{\alpha(1)}}(\hat{D}).
\]
If we define a value assgining map $v$ by $v(\hat{O}) = v_{\cx{\alpha(1)} \to \cdots \to\cx{\alpha(k)}}(\hat{O})$ if $\hat{O} \incx \cx{\alpha(k)}$, then $v$ satisfies FUNC partially, i.e., for $f^{j}$s ($j=1, \ldots, N$), but this contradicts the assumption for $\hat{D}$.
The root of this contradiction  comes from the assumption that $\hat{D} \incx \cx{\alpha(j)}, \forall j$.
Thus there must be a $j$ such that $\hat{D}$ is not stable in $\cx{\alpha(j)}$.

\section*{Acknowledgments} 
I thank Professor L. Accardi for useful comments and people of the Volterra Center for their warm hospitality.
I also thank my colleagues at Hokusei Gakuen University Junior College for their support, especially late Professor Reimei Kobayashi for his warm encouragement.

\end{document}